**Title:**

Rapid single-shot parity spin readout in a silicon double quantum dot with fidelity exceeding 99 %

**Author list:**

Kenta Takeda[1, *], Akito Noiri[1], Takashi Nakajima[1], Leon C. Camenzind[1], Takashi Kobayashi[2], Amir Sammak[3,4], Giordano Scappucci[3,5], and Seigo Tarucha[1, 2 *]

**Affiliations:**

1 RIKEN Center for Emergent Matter Science (CEMS), Wako, Japan.

2 RIKEN Center for Quantum Computing (RQC), Wako, Japan.

3 QuTech, Delft University of Technology, Delft, The Netherlands.

4 Netherlands Organisation for Applied Scientific Research (TNO), Delft, The Netherlands.

5 Kavli Institute of Nanoscience, Delft University of Technology, Delft, The Netherlands.

*Correspondence to: Kenta Takeda (kenta.takeda@riken.jp) or Seigo Tarucha (tarucha@riken.jp)

**Abstract**:

Silicon-based spin qubits offer a potential pathway toward realizing a scalable quantum computer owing to their compatibility with semiconductor manufacturing technologies. Recent experiments in this system have demonstrated crucial technologies, including high-fidelity quantum gates and multiqubit operation. However, the realization of a fault-tolerant quantum computer requires a high-fidelity spin measurement faster than decoherence. To address this challenge, we characterize and optimize the initialization and measurement procedures using the parity-mode Pauli spin blockade technique. Here, we demonstrate a rapid (with a duration of a few μs) and accurate (with >99% fidelity) parity spin measurement in a silicon double quantum dot. These results represent a significant step forward toward implementing measurement-based quantum error correction in silicon. (116 words)

**Main text**:

High-fidelity measurement of quantum states is crucial for the operation of a quantum computer. This is particularly important for quantum error correction protocols, which rely on feedback quantum gates that are based on syndrome measurements [1,2]. In order to implement these protocols effectively, the qubit measurements must be performed much faster than decoherence. While previous experiments with spin qubits in silicon have achieved significant milestones such as quantum non-demolition measurement [3,4], high-fidelity quantum gates [5–9], and multi-qubit control [10–12], single-shot measurement of single-spin states has generally been slow and with modest fidelity.

The readout of a single-spin state typically relies on spin-to-charge conversion techniques, such as energy-selective tunneling [13] or Pauli spin blockade (PSB) [14,15]. The energy-selective readout requires high bandwidth measurement with a long duration to detect stochastic tunneling events with various time scales. While this scheme has been employed for demonstrations of high-fidelity single-spin readout in the literature, the readout is typically orders of magnitude slower than phase coherence times [9,16]. On the other hand, the PSB readout utilizes spin-selective tunneling between two quantum dots. In this case, the signal results from a stationary difference of charge states, and charge discrimination is much easier. Here, we demonstrate that it is possible to perform a single-spin readout with a fidelity well above 99 % with the standard PSB mechanism. The spin readout can be performed within 3 us and is considerably faster than the average spin echo coherence times of around 100 us. We show coherent Rabi oscillations with visibility approaching 99.6 % (or a state preparation and measurement (SPAM) fidelity of 99.8 %), potentially compatible with measurement-based quantum error detection and correction protocols.

Figure 1a shows a scanning electron microscope image of our device. The overlapping aluminum gates [17] are fabricated on top of an isotopically enriched silicon/silicon-germanium wafer (heterostructure B in [18]). The sample is cooled down in a dilution refrigerator and an external magnetic field of 0.5 T is applied. In this work, we form two quantum dot spin qubits $Q_L$ and $Q_R$ under gates P3 and P4, respectively, while the left part of the device serves as an extended reservoir for $Q_L$. Gate B3 is used to control the tunnel coupling between the left reservoir and $Q_L$, and gate B4 is used to control the inter-dot tunnel coupling $t_c$. We monitor the occupancy of the double quantum dot by a nearby charge sensor quantum dot (indicated by the upper circle in Fig. 1a). The conductance of the charge sensor is measured by a radio-frequency reflectometry technique [19,20] with a bandwidth of approximately 10 MHz, which is limited by the quality factor of the tank circuit. In what follows, we use virtual gate voltages (vB3, vP3, vB4, vP4) to individually control quantum dot potential and tunnel coupling [21]. The single-qubit manipulation is realized by an electric-dipole spin resonance (EDSR) technique using a micromagnet [22]. The qubit manipulation is performed at the charge symmetry point, where the qubit energy is first-order insensitive to detuning charge noise [23,24]. The residual exchange coupling is measured to be 21 kHz (see Supplementary Fig. 1).

We use the four-electron charge states (4,0) and (3,1) for the PSB readout, where $(n_L, n_R)$ denotes the number of electrons in the left and right quantum dots. For the (4,0) configuration, the first excited state is energetically split by the orbital splitting rather than the valley splitting [11]. This results in a larger readout window since the orbital excitation energy (400 μeV for the four-electron state) is larger than the valley splitting (100-150 μeV in this device). In Fig. 1b, we measure a charge stability diagram as a function of vP3 and vP4. We rapidly ramp vP4 from the (3,1) configuration towards the (4,0) configuration, causing the (3,1) triplet state to latch in the PSB regime. We choose an integration time of 10 μs per point to ensure that relaxation of the triplet state is not significant. The PSB regime appears as an extended trapezoidal region showing (3,1) charge signal within the (4,0) configuration (see inside the white rectangle in Fig. 1b).

We begin with characterizing and optimizing the initialization protocol. Our initialization procedure starts by waiting near the (3,0)-(4,0) boundary where energy-selective tunneling to the (4,0) singlet state S(4,0) occurs. The waiting time is set to 5 μs, and we pulse vB3 to increase the initialization rate. The singlet-triplet excitation energy being much larger than the thermal broadening (~300 μeV vs. ~4 μeV) results in negligible population of higher excited states. We subsequently apply a voltage ramp to separate S(4,0) to $|\uparrow\downarrow\rangle$ by a rapid adiabatic passage as illustrated by the dashed black arrow in Fig. 2a. For spin qubits with micromagnet, the primary concern in this procedure is the adiabatic transition through the $S - T_-$ anti-crossing. The size of this anti-crossing is determined by the magnetic field gradient and $t_c$, while only the latter is well controllable through gate voltages. To mitigate the adiabatic transition, we use a ramp procedure shown in Fig. 2b. Here the detuning is defined as vP4−vP3 using virtual gate voltages. We first pulse the system to the (3,1) regime close to the interdot boundary and increase $t_c$ by opening vB4. The concept is to make $t_c$ sufficiently small around the PSB regime and the $S - T_-$ anti-crossing, and large enough at the inter-dot transition, while ensuring a simple pulse shape for routine device calibration. The inter-dot coupling is then decreased during the ramp toward point O to adiabatically decrease the exchange coupling (see Supplementary Fig. 2). After the qubit manipulation at O, we can perform a projective measurement by a reverse pulse sequence to M and charge discrimination. To optimize vB4 for the ramp pulse, we measure the triplet return probability as a function of vB4 at point R in Fig 2b. Here, too large vB4 results in adiabatic transition to $T_-$, while too small vB4 results in diabatic inter-dot transition causing the (3,1) charge state latching signal. For the best parameter used in the experiment, we obtain roughly 0.2% triplet probability after pulsing to the (3,1) state (Fig. 2c). Figure 2d shows a comparison of the readout outcomes for the states right after S(4,0) initialization and after the ramp. We see that the S(4,0) initialization error is negligible, and the ramp to the (3,1) state causes an increased error counts.

In the PSB regime, there exist three possible spin-blocked states: $T_-$, $T_+$, and $|\downarrow\uparrow\rangle$. Among these three states, $|\downarrow\uparrow\rangle$ quickly mixes with the unblocked $|\uparrow\downarrow\rangle$ state via the micro-magnet field gradient ($T_1 \sim 100$ ns). As a result, in practice, only the spin-polarized triplet states $T_-$ and $T_+$ are blocked, and the readout is the so-called parity readout. In Fig. 3a, b, we show measurement of the $T_1$ relaxation times of triplet states $T_-$ and $T_+$. These two triplet states are prepared by applying a $\pi$ rotation to $|\uparrow\downarrow\rangle$. We obtain $T_1 = 18.2 \pm 0.2$ ($3.268 \pm 0.001$) ms for $T_-$ ($T_+$) by fitting the data to exponential decay. These long $T_1$ values are achieved by reducing $t_c$ during the readout (see Supplementary Fig. 3). While the field-gradient components causing the spin relaxation are similar in magnitude for these two cases, there are several possible mechanisms that cause this difference. For example, the respective detuning parameters for the relevant spin-flipping anti-crossings ($S(4,0) - T_\pm$) are different for these two states, and the associated energy differences being compensated by phonons, charge noise, and/or Johnson-Nyquist noise also differ [25,26]. In addition, for the case of $T_+$, sequential relaxation via the unpolarized spin states may occur. Pinpointing the exact mechanisms limiting these $T_1$ values, including additional experiments on parametric dependence and related theory, is a subject for future study. Nonetheless, these are both long enough for high-fidelity parity spin measurement.

Next, we move on to characterizing the signal-to-noise ratio (SNR), which is another important metric for spin readout. Figure 3c shows the results of charge discrimination after preparing an equal superposition of $|\uparrow\downarrow\rangle$ and $T_-$. Two peaks in each histogram correspond to singlet and triplet readout outcomes. If the integration time $t_{\text{int}}$ is too short, the overlap of two Gaussian peaks causes a substantial infidelity (upper panel in Fig. 3c). However, if $t_{\text{int}}$ is too long, the relaxation of excited (3,1) state reduces the readout fidelity. The optimal $t_{\text{int}}$ to minimize readout infidelity is obtained by balancing these two factors. Figure 3d plots the readout infidelity taking into account both $T_1$ and SNR. For both $T_-$ and $T_+$, we obtain measurement infidelities below 0.1% with $t_{\text{int}} = 2$ μs. For experiments requiring measurement-based feedback, it might be important to reduce $t_{\text{int}}$ further to minimize the dephasing of idling qubits. For reduced $t_{\text{int}} = 0.5$ μs, the infidelities are higher but remain below 1 %.

Finally, we assess the SPAM fidelity by measuring coherent Rabi oscillations. Throughout these experiments, we use $t_{\text{int}} = 3$ μs. In Fig. 4a, we measure two resonance peaks for $Q_L$ and $Q_R$, with a difference in resonance frequencies corresponding to a Zeeman energy difference of $\Delta E_z = 85.6$ MHz. The Rabi frequency is chosen so that the amplitude of the off-resonant drive of the idle qubit is negligibly small. We measure Rabi oscillations by varying the microwave pulse time $t_p$ (Fig. 4b, c). We find that the Rabi oscillation decay is negligible, and the oscillation visibilities are well above 99% for both qubits ($99.6 \pm 0.2\%$ for $T_-$ and $99.4 \pm 0.2\%$ for $T_+$). For the $T_-$ readout, the 0.2% initialization error due to the adiabatic $S - T_-$ transition limits the visibility. For the $T_+$ readout, the $T_1 \sim 3$ ms accounts for $\sim 0.1\%$ of the additional visibility loss as compared with the case for $T_-$. We speculate that the additional 0.1% of infidelity comes from a larger adiabatic transition probability through the $S - T_+$ anti-crossing where vB4 is larger than the $S - T_-$ anti-crossing, as seen in the pulse shape of Fig. 2b. A more complex pulse to reduce $t_c$ when crossing both degeneracy points could circumvent this problem.

For implementing measurement-based quantum protocols that require phase coherence, the total duration of readout is also crucial. More explicitly, measurement has to be faster than relevant $T_2$, not only $T_1$. Our spin qubits have $T_2^* \sim 8$ μs (see Supplementary Fig. 4), and these inhomogeneous phase coherence times can be extended to above 100 μs using a single refocusing $\pi$ pulse (Fig. 4b). On the other hand, the overall duration of our measurement procedure can be as short as 2.4 μs (0.13 μs ramp time + 0.25 μs reflectometry signal rise time + 2 μs integration time). When comparing this duration to the typical loss of coherence expected from the Hahn echo decay of $Q_L$ and $Q_R$, we expect less than a 1 % loss of phase coherence during the measurement (0.1% (0.5%) for $Q_L$ ($Q_R$)). While a practical way to combine the spin echo pulse and the spin readout scheme remains to be explored, our result suggests that, with an appropriate device and experimental configuration, it is possible to perform a single-spin qubit readout precisely and fast enough while maintaining phase information.

In conclusion, we have demonstrated high-fidelity parity PSB readout for electron spin qubits in a silicon double quantum dot. The readout fidelities compare favorably to the best values in other spin qubit platforms [16,27,28]. The visibility of Rabi oscillation is mainly limited by spin state mapping, which could be improved by further optimization of pulse shape. The readout duration is much shorter than the spin echo dephasing time. We note that the implemented parity spin readout does not preserve the original eigenstates unlike the standard ZZ parity measurement, and we need a designated measurement ancilla quantum dot to readout a single-spin state. It may suggest that optimizing only the $T_-$ readout is sufficient for applications where we need single-spin projective measurement. Overall, our results suggest that it is possible to perform a single-spin readout in a time scale where the phase coherence of idling qubits is well preserved, and it may open the door toward realizing measurement-based quantum error detection or correction protocols in silicon.

**Data availability:**
The data used to produce the figures in this work is available from the Zenodo repository at {URL}.

**Acknowledgements**:
We acknowledge Xin Liu and Reiko Kuroda for assistance with sample fabrication. This work was supported financially by JST Moonshot R&D Grant Number JPMJMS226B, and JSPS KAKENHI grant Nos. 18H01819, 20H00237, and 23H05455. T.N. acknowledges support from JST PRESTO Grant Number JPMJPR2017. L.C.C. acknowledges support from Swiss NSF Mobility Fellowship No. P2BSP2 200127.

**Figures:**

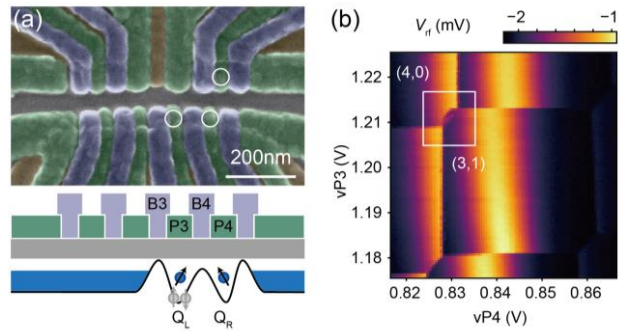

Figure 1 (a) False-colored scanning electron microscope image of the device. The green, purple, and brown gates correspond to plunger/accumulation, barrier, and screening gates, respectively. The lower channel is used for defining the right and left quantum dots under the plunger gates P3 and P4 (lower panel). A micromagnet is fabricated on top of the fine gate stack. The right quantum dot in the upper channel is used as a charge sensor quantum dot. The right bath for the charge sensor is connected to a radio frequency tank circuit. The white circles indicate where quantum dots are formed. (b) Charge stability diagram around the (4,0) - (3,1) charge degeneracy point. $V_{\rm rf}$ is the reflectometry signal from the charge sensor. The vP4 voltage sweep is performed from the right to left so that the triplet (3,1) state is latched in the spin blockade regime.

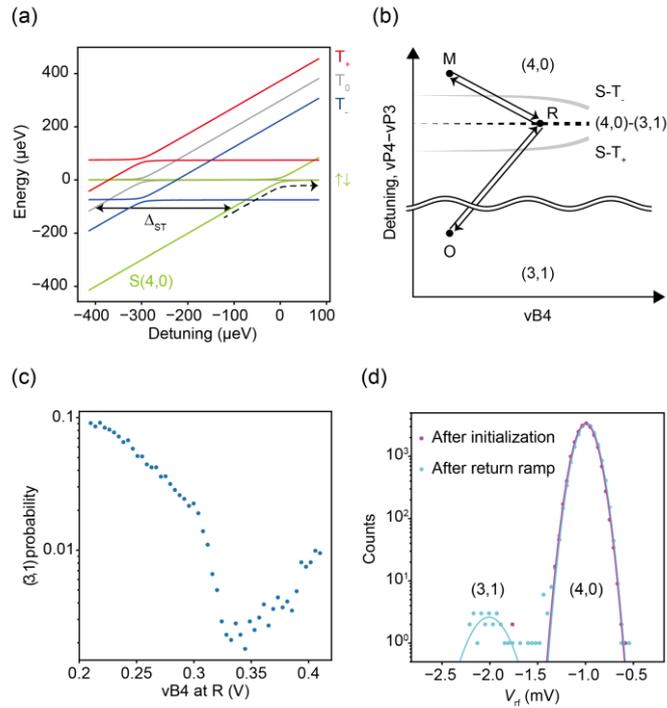

Figure 2. Initialization procedure. (a) Energy diagram of the two-electron states around the inter-dot charge transition. The zero detuning point is defined as the S(4,0)-S(3,1) degeneracy point. The dashed line represents the ideal rapid adiabatic passage for initializing $|\uparrow\downarrow\rangle$. We use parameters $\Delta_{ST} = 400$ μeV, average Zeeman energy of 18 GHz, Zeeman energy difference of 100 MHz, and $t_c = 2$ GHz for the illustration. (b) Schematic of initialization pulse trajectory in the vB4-detuning plane. The durations are 15 ns (ramp time from M to R), 15 ns (waiting time at R), and 100 ns (ramp time from R to O). The position of $S - T_\pm$ crossings is illustrated for $t_c$ changing exponentially as a function of vB4 (solid gray lines). (c) Measurement of triplet return probability as a function of the virtual barrier gate voltage vB4 at R. (d) Histograms of the charge sensor signal for 10,000 shots of measurement before and after the ramp to the (3,1) charge state. The solid lines show a fitting to a Gaussian function. The (3,1) counts for the configuration right after initialization is too small to fit.

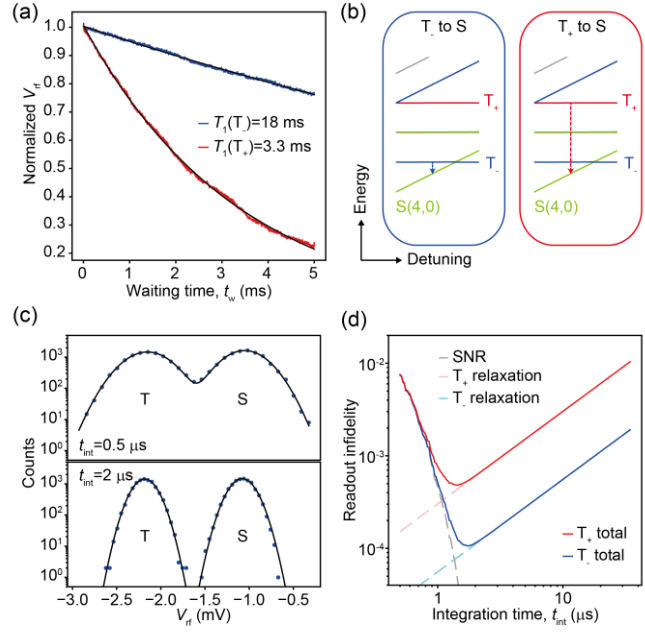

Figure 3 Relaxation, SNR, and readout fidelity. (a) Relaxation times. The blue (red) trace shows the measurement result for the initial state $T_-$ ($T_+$). The data is obtained by normalizing the measured $V_{rf}$ so that $V_{rf}(t_w = 0) = 1$ and $V_{rf}(t_w = \infty) = V_{rf}(S(4,0)) = 0$. The S(4,0) signal is obtained by measuring $V_{rf}$ at M directly after the initialization. The solid black lines are the exponential fit curves $V_{rf}(t_w) = V_0 \exp(-t_w/T_1)$, where $V_0 \sim 1$ is the fitting parameter to account for the signal contrast. (b) Zoom up of the energy diagram plotted with relevant relaxation channels (left panel: S to $T_-$, right panel: S to $T_+$). (c) Histograms for two different measurement integration times $t_{int} = 0.5$ μs and 2 μs. The blue points show histogram counts for measured data. The solid black lines show fitting to a double Gaussian function. (d) Infidelities as a function of $t_{int}$. The SNR-inferred infidelity is calculated from the overlap of two Gaussian peaks, and the $T_1$-inferred infidelity is defined as $1 - \exp(-t_{int}/T_1)$.

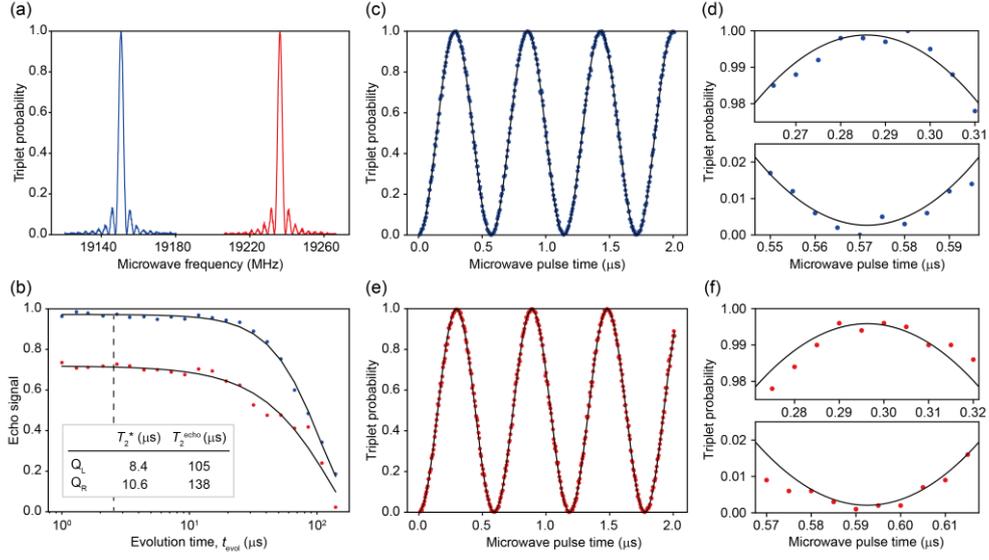

Figure 4. Spin qubit control and SPAM fidelity. (a) EDSR spectra of two spin qubits. The blue (red) trace shows the measured data of $Q_L$ ($Q_R$). The resonance frequencies are 19.1506 GHz for $Q_L$ and 19.2362 GHz for $Q_R$. (b) Spin echo measurements. The evolution time refers to the total duration of waiting time between microwave pulses. The blue (red) points show the measured data of $Q_L$ ($Q_R$). The black lines show the fit with exponential decay $P(t_{evol}) = V\exp(-(t_{evol}/T_2^{echo})^\gamma)$ from which we extract $T_2^{echo} = 105 \pm 1\,(138 \pm 8)$ μs, $\gamma = 1.89 \pm 0.07\,(1.3 \pm 0.1)$, and $V = 0.972 \pm 0.002\,(0.969 \pm 0.002)$ for $Q_L$ ($Q_R$). The data points and fit curve for $Q_R$ are offset by $-0.25$ for clarity. The non-ideal visibility is due to the imperfect calibration of microwave pulses. The dashed black line represents $t_{evol} = 2.4$ μs. (c) Rabi oscillation of $Q_L$. The blue points are measured data, and the black line is a sinusoidal fit. (d) Zoom up of a peak and dip in c. (e,f) measurements similar to c and d, but for $Q_R$.

**Supplementary Information for "Rapid single-shot parity spin readout in a silicon double quantum dot with fidelity exceeding 99 %"**

**Setup**

The sample is cooled down in a dilution refrigerator. The electron temperature is estimated to be around 45 mK by observation of the thermal broadening of the charge transition line [29]. The charge sensor quantum dot is connected to an LC tank circuit with an inductance of 1.2 µH and a resonance frequency of 181 MHz. The rf carrier signal is generated by an internal AWG of a Keysight M3302A module running at 500MSa/s. The signal is attenuated and applied to the device via a directional coupler mounted at the mixing chamber plate of the dilution refrigerator. The reflected signal is first amplified by a Cosmic microwave CITLF2 cryogenic amplifier and further amplified by room temperature amplifiers. The amplified signal is then digitized by the M3302A module at a sampling rate of 500MSa/s. Its internal FPGA is used to demodulate the signal to the baseband voltage $V_{rf}$. The gate electrodes P3, P4, B3, and B4 are connected to a Keysight M3202A AWG module running at 1GSa/s via cryogenic bias-tees. The microwave signal is generated by two Keysight E8267D vector signal generators that are IQ modulated by a Keysight M3202A module. Unless noted, we collect 1,000 single-shot measurement outcomes to obtain a triplet return probability.

**Residual exchange coupling**

The residual exchange coupling is measured by a spin-echo like sequence shown in Supplementary Fig. 1a [30]. This sequence cancels out single-qubit phase accumulation but does not affect two-qubit conditional phase. For the condition used for single-qubit gates (vB4=0.28), we obtain a residual exchange coupling of 21 kHz by fitting the data.

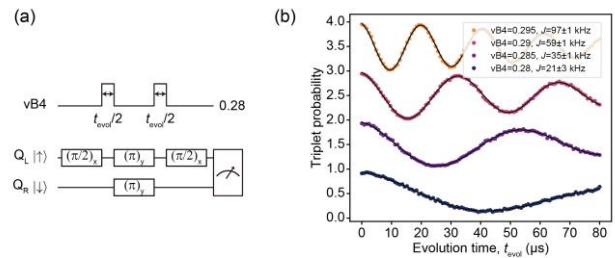

Supplementary Fig. 1. Measurement of residual exchange coupling. **a**, Pulse sequence used for the experiment. **b**, Measured decoupled CZ oscillation for various barrier gate voltages. The errors represent one sigma from the mean.

**Controlled rotation for checking adiabatic spin state initialization**

In the main text, we design our pulse so that it selectively initializes a $|\downarrow\uparrow\rangle$ state by pulsing the exchange coupling slowly enough compared with $\Delta E_z$. However, just by observing the single-qubit rotations demonstrated in Fig. 4, it is impossible to check if there is a substantial Landau-Zener transition into the $|\uparrow\downarrow\rangle$ state. To infer this, we perform a controlled-rotation experiment. Here, we measure the left qubit resonance peaks at a charge symmetry point with an exchange coupling of about 25 MHz. Indeed, we see the peak position depends on the state of the control qubit ($Q_R$) (Supplementary Fig. 2a). If we have imperfect initialization to the $|\uparrow\downarrow\rangle$ state, there is a finite population of the $|\downarrow\uparrow\rangle$ state. For the Z-CROT signal (red trace in Supplementary Fig. 2a), it results in an additional peak at the resonance condition for CROT. In our experiment, we do not see a clear peak structure in our experiment (Supplementary Fig. 2b). For the number of samplings used in this experiment ($N = 2,000$), we are not able to distinguish a resonance peak from the noise expected from binomial distribution (Supplementary Fig. 2c, standard deviation = 0.0017). Therefore, we expect the residual population in $|\downarrow\uparrow\rangle$ is at most comparable to this standard deviation. To measure it with higher precision, it is necessary to increase $N$ significantly, but the limited stability of the device used in the experiment did not allow for such a time-consuming experiment. Nonetheless, this initialization error would not be a practical limitation since it can be suppressed by increasing the ramp time.

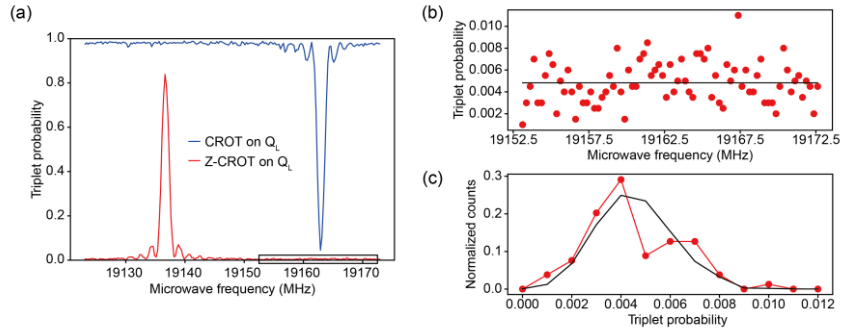

Supplementary Fig. 2. Controlled rotation. **a**, EDSR spectra for two control qubit states $Q_R=|\downarrow\rangle$ (red) and $Q_R=|\uparrow\rangle$(blue). The imperfect visibility is mostly due to the miscalibration of microwave pulse amplitude or length. We collect 2,000 single-shot data per one data point. **b**, Zoom up around the CROT resonance frequency $f = 19162.5$ MHz. The red points are measured data and the black line shows the average of the plotted data (0.004829). This average value is somewhat higher than the one measured in Fig.4 due to imperfect calibration. **c**, Comparison of the histogram of measured data (red points) and the resampled data from a binomial distribution with an expectation value of 0.004829 (black curve).

**Inter-dot tunnel coupling dependence of $T_1$ during spin readout**

Supplementary Fig. 3 shows $T_1$ measured as a function of vB4 during spin readout. The direct measurement of $t_c$ is difficult for the range of vB4 used in this measurement. As a reference, supplementary Fig. 3b shows extrapolation from measurements where $t_c$ is significantly larger than the thermal broadening.

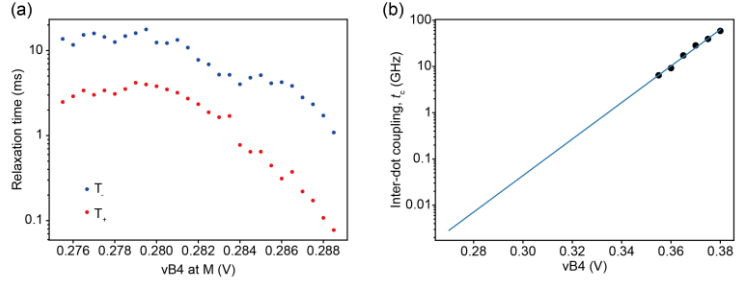

Supplementary Fig. 3, **a**, $T_1$ of T_ and T_+ measured as a function of vB4. **b**, Inter-dot coupling $t_c$ measured as vB4. The black points are obtained by fitting the broadening of inter-dot transition [29]. The blue line is a fit to an exponential function.

**Ramsey interferometry**

We perform a standard Ramsey interferometry to measure the $T_2^*$ of qubits $Q_L$ and $Q_R$. The results are shown in Supplementary Fig. 4. We fit the data with an exponential decay $P(t_{evol}) = A\exp\left(-\left(\frac{t_{evol}}{T_2^*}\right)^\alpha\right)\cos(2\pi ft) + B$. From the fit, we obtain $A = 0.494 \pm 0.005$ ($0.495 \pm 0.006$), $B = 0.500 \pm 0.001$ ($0.493 \pm 0.002$), $\alpha = 2.04 \pm 0.08$ ($1.8 \pm 0.1$), and $T_2^* = 8.36 \pm 0.08$ ($10.6 \pm 0.2$) μs for $Q_L$ ($Q_R$). The data acquisition time is 28.2 sec for both data.

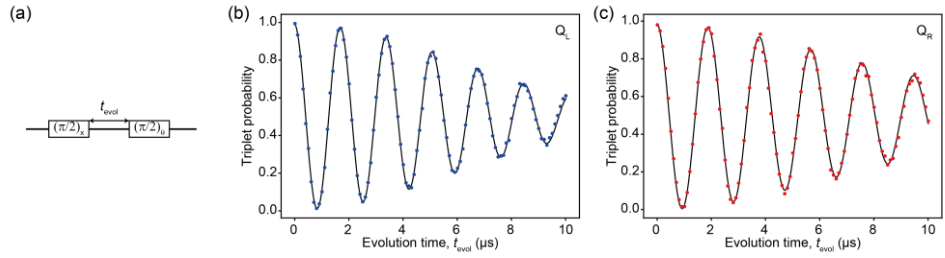

Supplementary Fig. 4. Ramsey interferometry. **a**, Microwave control pulse used for the experiment. To facilitate the curve fitting, the phase of the second $\pi/2$ pulse is varied as $\theta = 2\pi t_{evol} \times 0.5$ (μs · MHz) so that the resulting probability oscillates at a frequency of about 0.5 MHz. **b**, **c**, Measurement result for $Q_L$ ($Q_R$). The blue (red) points represent the data, and the black solid line shows a fit.